\documentclass{svjour2}
\usepackage{graphicx}  
\usepackage{latexsym}
\usepackage{amssymb}

\def\beq{\begin{equation}}
\def\eeq{\end{equation}}
\def\rmd{{\rm d}} 

\journalname{General Relativity and Gravitation}

\begin{document}

\title{Refraction index analysis of light propagation in a colliding gravitational wave spacetime}

\author{Donato Bini \and
        Andrea Geralico \and
        Maria Haney         
}

\institute{
              Donato Bini 
              \at
              Istituto per le Applicazioni del Calcolo ``M. Picone,'' CNR, I-00161 Rome, Italy\\
              ICRA, University of Rome ``La Sapienza,'' I--00185 Rome, Italy\\
              INFN - Sezione di Firenze, Polo Scientifico, Via Sansone 1, I--50019, Sesto Fiorentino (FI), Italy\\
              \email{binid@icra.it} 
\and
              Andrea Geralico 
              \at
              Physics Department and ICRA, University of Rome ``La Sapienza,'' I--00185 Rome, Italy\\
              \email{geralico@icra.it}   
\and
              Maria Haney 
              \at
              Physics Department and ICRA, University of Rome ``La Sapienza,'' I--00185 Rome, Italy\\
              \email{haney@icra.it}          
}

\date{Received: date / Accepted: date / Version: date }

\maketitle

\begin{abstract}
The optical medium analogy of a given spacetime was developed decades ago and has since then been widely applied to different gravitational contexts. Here we consider the case of 
a colliding gravitational wave spacetime, generalizing previous results concerning single gravitational pulses. Given the complexity of the nonlinear interaction of two gravitational waves in the framework of general relativity, typically leading to the formation of either horizons or singularities, the optical medium analogy proves helpful  to simply capture some interesting effects  of photon  
propagation.

\end{abstract}

\section{Introduction}

The electromagnetic field equations in a curved spacetime can be formally cast into the form of Maxwell's equations in a flat spacetime but in the presence of an \lq\lq effective medium'' with the dielectric permittivity and magnetic permeability tensors properly related to the spacetime metric (see, e.g., Refs. \cite{pleb,skrotskii,fdf1,mas73}).
Such a material medium is endowed with a specific refraction index (or more than one, if it is anisotropic), which takes over the role of the non-Minkowskian terms of the background metric.
For instance, electromagnetic waves propagate through the Schwarzschild spacetime as though in a uniaxial anisotropic
medium at rest in an inertial frame, whereas the Kerr geometry acts as a biaxial anisotropic medium, except on the axis of symmetry \cite{hanni}.
Such an approach operationally consists in projecting the electromagnetic field onto a flat spacetime and suitably modifying the sources.
It has been largely adopted in the literature to study different aspects of propagation of electromagnetic waves in a gravitational field as well as scattering processes by compact objects \cite{mas73,bal,mas74,mas75}, in spite of its limited applicability.
In fact, the constitutive equations specifying the medium are not strictly covariant, as already pointed out by Plebanski, and the analogy with macroscopic electrodynamics makes sense only if the coordinates used are Cartesian-like. 

The analogy between electromagnetism in a curved spacetime and in a material medium was worked out also by Landau and Lifshitz \cite{LL}.
Instead of dealing with projected electric and magnetic field densities as well as modified sources, they defined the electromagnetic fields in a curved 3-space and used the actual sources.
The curved background formalism developed by Landau and Lifshitz thus gives a global view of electromagnetic fields with respect to a local static observer, but it is limited to those regions of spacetime where static observers do exist.
The material medium analogy presented by Plebanski gives, instead, the viewpoint of an observer at infinity due to the projection onto a flat background.
 
In the high frequency approximation, i.e., when the wavelength of electromagnetic waves is much smaller than the characteristic scale of the gravitational field, electromagnetic waves propagate along the null geodesics of the background spacetime.
In fact, the wave equations are reduced to the Hamilton-Jacobi equation for the eikonal function, whose characteristics are just the null geodesics of the spacetime metric, and to transport equations for the slowly varying amplitude of the wave.
This is the well known geometrical optics limit of Maxwell's equations in a gravitational field (see, e.g., Ref. \cite{mas87} and references therein).

In this paper we apply the optical medium analogy of a given background recalled above to an exact solution of the Einstein's field equations belonging to the class of two colliding gravitational plane wave spacetimes.
Such solutions represent some of the simplest exact dynamical spacetimes, providing clear examples of highly nonlinear behavior in general relativity.
In fact, as a result of the collision process, the focusing effects of each exact plane wave lead to mutual focusing, which yields the formation of either a spacetime singularity or a nonsingular Killing-Cauchy horizon at the focusing points of the two waves \cite{szek70,kha-pe,griff}.
These waves have then been used both in classical general relativity to test some conjectures on the stability of Cauchy horizons \cite{ori,yurt}, and in string theory to investigate classical and quantum string behavior in strong gravitational fields \cite{vega,nunez}.
Furthermore, exact plane waves may play a role in the study of the strong time dependent gravitational fields produced in the collision of black holes \cite{fpv}, or even represent travelling waves on strongly gravitating cosmic strings \cite{garf}.

To the best of our knowledge, in the literature so far only the optical medium analogy of single and weak gravitational wave pulses has been studied.
Mashhoon and Grishchuk \cite{masgris80} proved that electromagnetic phenomena in the background of a weak gravitational wave radiation field propagate with a frequency which is simply related to that of the background solution.
The case of exact plane waves is expected to be very different from their linearized counterparts, which have no focusing points and admit a globally hyperbolic spacetime structure.
Recently we have worked out the optical medium analogy of a radiation field sandwiched between two flat regions, considering a scenario where the radiation field is represented either by an exact gravitational wave or by an exact electromagnetic wave in general relativity \cite{gwem_epl}.
In the present paper we concentrate on the case of two exact colliding plane gravitational waves, exploring in particular the collision region associated with the nonlinear interaction of the waves.
As recalled above the collision process of two single gravitational waves can be associated with two possible scenarios: either the formation of a Killing-Cauchy horizon or that of a spacetime singularity.
There are a variety of possible solutions which exhibit one of the two above mentioned features (see, e.g., Ref. \cite{griff}). 
We consider here the one found by Ferrari and Iba\~nez \cite{Fe-Ib,Fe-Ibbis,Fe-Ib2} in the 1980s. It has the twofold advantage of being mathematically simple and allowing one to switch easily between the horizon-forming and singularity-developing solution by a sign change in certain metric functions (which are in turn summarizable with a sign indicator).
We are interested in studying the properties of the optical medium analog of such a colliding gravitational wave spacetime.
We expect that the nonlinear wave interaction leads to a significant modification of the single wave picture, especially in connection with the presence of singular structures which usually form during the collision process. 

Units are such that $G=1$ and $c=1$, $G$ being the Newtonian constant and $c$ the speed of light in vacuum. Greek indices run from $0$ to $3$ and latin indices run from $1$ to $3$. The metric signature is chosen to be $+2$.

\section{The optical medium analogy}

Consider an arbitrary gravitational field described by the line element $ds^2=g_{\alpha\beta}dx^\alpha dx^\beta$ written in Cartesian-like coordinates $x^\alpha=(t,x,y,z)$. 
An electromagnetic field in that background can be thought of as propagating in flat spacetime but in the presence of a medium whose properties are determined by conformally invariant quantities constructed from the metric tensor. 
In fact, the covariant Faraday tensor  $F_{\mu\nu}$ and its rescaled contravariant counterpart $\sqrt{-g}F^{\mu\nu}$ can be decomposed as $F_{\mu\nu}\to ({\mathbf E},{\mathbf B})$ and $\sqrt{-g}F^{\mu\nu}\to (-{\mathbf D},{\mathbf H})$ to yield the usual Maxwell's equations in a medium \cite{pleb,skrotskii,fdf1,mas73}
\begin{eqnarray}
\label{MWEQ}
 \nabla \cdot {\mathbf B}=&0\,, \qquad\nabla \times {\mathbf E}&=-\partial_t {\mathbf B}\,,\nonumber\\
 \nabla \cdot {\mathbf D}=&4\pi\rho\,, \quad\nabla \times {\mathbf H}&=\partial_t {\mathbf D}+4\pi {\mathbf J}\,,
\end{eqnarray}
where the current vector $J_\mu \to (\rho , {\mathbf J})$ satisfies the conservation law
\beq
\partial_t \rho+ \nabla \cdot {\mathbf J}=0\,.
\eeq
The above set of equations is completed by the constitutive relations
\beq
\label{diel}
D_a=\epsilon_{ab}E_b -({\mathbf M}\times {\mathbf H})_a\,,\qquad 
B_a=\mu_{ab}H_b+({\mathbf M}\times {\mathbf E})_a\,,
\eeq
where
\beq
\label{epsdef}
\epsilon_{ab}=\mu_{ab}=-\sqrt{-g}\, \frac{g^{ab}}{g_{tt}}\, 
\eeq
play the role of  electric and magnetic permeability tensors and 
\beq
M_a=-\frac{g_{ta}}{g_{tt}}\
\eeq
is a vector field associated with rotations of the reference frame.
The effective material medium is in general anisotropic and has no birefringence, due to the proportionality between polarization tensors.
Furthermore,  the conformal invariance of Maxwell's equations is reflected, in the present formulation, by the independence of the dielectric tensors $\epsilon_{ab}$ and $\mu_{ab}$ from a conformal factor in the metric components, a property also shared by the \lq\lq spatial" vector ${\mathbf M}$.

In the absence of currents, if the metric tensor varies in space and time only slightly with respect to the wavelength and the period of the wave, one may look for solutions of the form
\beq
\label{toreplace}
{\mathbf E}={\mathbf E}_0 e^{i(n {\mathbf k}\cdot {\mathbf x}-\omega t)}\,,\quad  {\mathbf E}_0=const.\,,\quad |{\mathbf k}|=\omega\,,
\eeq
where $n$ is an effective refraction index.
Similar expressions hold for ${\mathbf B}$, ${\mathbf H}$ and ${\mathbf D}$. 
Maxwell's equations (\ref{MWEQ}) then imply the following generalized Fresnel equation \cite{skrotskii}
\beq
\label{fresneleq}
n^2\epsilon_{{\mathbf e}{\mathbf e}}-2n\epsilon_{{\mathbf e}{\mathbf M}}+\epsilon_{{\mathbf M}{\mathbf M}}-{\rm det}(\epsilon)=0\,,
\eeq
where ${\mathbf e} = {\mathbf k}/\omega$ is the spatial unit vector of the photon direction and the compact notation $X_{ab}A^a B^b = X_{{\mathbf A}{\mathbf B}}$ has been introduced for contraction of a generic matrix $X_{ab}$ with vectors $A_a$ and $B_b$. The above equation (\ref{fresneleq}) gives the relation between the effective refraction index of the medium, the components of the polarization tensors and the direction of propagation of the electromagnetic wave. The solution for the refraction index is thus given by
\beq
\label{n_fin}
n = \frac{1}{\epsilon_{{\mathbf e}{\mathbf e}}}\left[\epsilon_{{\mathbf e}{\mathbf M}}+\sqrt{{\rm det}(\epsilon)\left[\epsilon_{{\mathbf e}{\mathbf e}}-(\epsilon^{-1})_{{\mathbf c}{\mathbf c}}\right]}\right]\,,
\eeq
where ${\mathbf c} = {\mathbf M} \times {\mathbf e}$.
In the special case $M_a=0$, the above expression (\ref{n_fin}) simplifies to
\beq
\label{refrac}
n = \sqrt{\frac{{\rm det}(\epsilon)}{\epsilon_{{\mathbf e}{\mathbf e}}}}\,.
\eeq

\section{Colliding gravitational wave spacetime}

Exact solutions of the Einstein equations representing colliding gravitational plane waves have been discussed extensively in the literature \cite{griff}.
The spacetime geometry associated with two colliding gravitational plane waves is characterized by the presence of either a spacetime singularity \cite{kha-pe} or a Killing-Cauchy horizon as a result of the nonlinear wave interaction \cite{Fe-Ib,Fe-Ibbis,Fe-Ib2}. 
In general, such spacetimes contain four regions: a flat spacetime region (Petrov type-O), representing the initial situation before the passage of the two oppositely directed plane waves, two Petrov type-N regions, corresponding to the single waves before the interaction, and an interaction region, generally
of Petrov type-I. Two commuting spacelike Killing vectors are always present, associated with the plane symmetry assumed for the two colliding waves.
The spacetime describing the collision region is most generally characterized by the line element \cite{szekeres}
\begin{eqnarray}
\rmd s^2&=&-e^{M}(\rmd t^2-\rmd z^2)+e^{-U}\left[-2\sinh W \rmd x\, \rmd y\right.\nonumber\\
&&+\cosh W\left(e^V\rmd x^2+e^{-V}\rmd y^2\right)\left.\right]\,,
\end{eqnarray}
where all metric functions depend on $t$ and $z$.

The electric and magnetic permeability tensors (\ref{epsdef}) of the corresponding equivalent medium are then given by
\beq
\epsilon_{ab}=\mu_{ab}=\left(
\begin{array}{ccc}
e^{-V}\cosh W & \sinh W & 0\cr
\sinh W & e^{V}\cosh W & 0\cr
0& 0 & e^{-(M+U)}
\end{array}
\right)\,,
\eeq
whereas the rotation vector $M_a=0$ vanishes identically in this case.

For a collinear polarization of the plane waves, the metric function $W$ can be gauged to be zero, so that the metric as well as the resulting electric and magnetic permeability tensors given above are diagonal.
One possible solution exhibiting these properties is the Ferrari-Iba\~nez metric considered below; it shares the most peculiar features of more general solutions of the same kind.

\subsection{The degenerate Ferrari-Iba\~nez metric}

\begin{figure}
\begin{center}
\includegraphics[scale=.9]{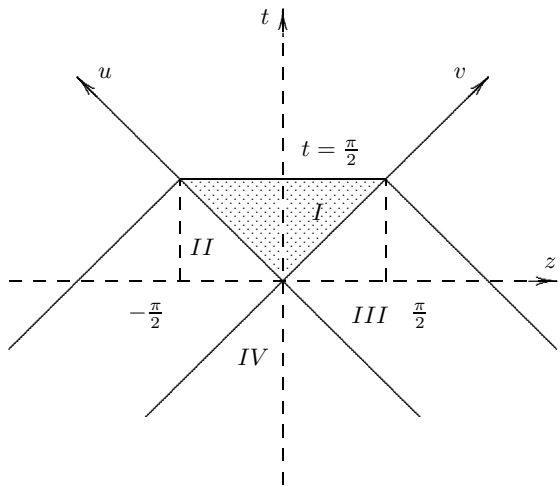} 
\end{center}
\caption{The null coordinates $(u,v)$ and the different regions they induce.
}
\label{fig:1}
\end{figure}

Ferrari and Iba\~nez \cite{Fe-Ib,Fe-Ibbis,Fe-Ib2} found a type-D solution of the Einstein equations that can be interpreted as describing the collision of two linearly polarized gravitational plane waves propagating along a common direction (the $z$-axis) in opposite senses and 
developing either a non-singular Killing-Cauchy horizon or a spacetime singularity upon collision. 
The corresponding line element in the interaction region  (hereafter referred to as Region I) can be written as
\beq
\label{met1}
\rmd s_{I}^2=- F_+^2(t)(\rmd t^2-\rmd z^2)+\frac{F_-(t)}{F_+(t)}\rmd x^2+ \cos^2zF_+^2(t)\rmd y^2\,,
\eeq
where 
\beq
F_\pm(t)=1\pm\sigma\sin t\,,\qquad \sigma=\pm 1\,,
\eeq
with $\sigma=1$ corresponding to a horizon-forming solution at time the $t=\pi/2$, whereas $\sigma=-1$ denotes a singularity-developing one.
The interaction region where this form of the metric is valid is depicted in the $t - z$ plane by a triangle whose vertex represents the initial event of collision and can be identified with the origin of the coordinate system; the horizon/singularity is mapped onto the base of the shaded triangle in Fig.~\ref{fig:1}. 
Therefore, Region I corresponds to the region $-t\leq z\leq t$, $0\leq t\leq \pi/2$. The instant of collision is $t=0$, while $t=\pi/2$ is the instant when the horizon/singularity is created.

If $t<0$ the two waves are traveling one against the other in the $z-$direction. They are single plane waves propagating
in flat spacetime.
In order to extend the metric from the interaction region to the remaining parts of the spacetime representing the single wave zones and the flat spacetime zone before the arrival of waves, one must first introduce the null coordinates
\beq
\label{eq:uv_trasf}
u=\frac{t-z}{2}\,, \quad v=\frac{t+z}{2} \qquad\Longleftrightarrow\qquad t=u+v\,, \quad z=v-u\,,
\eeq
in terms of which the metric (\ref{met1}) takes the form
\begin{eqnarray}
\label{met2}
\rmd s_{I}^2&=&-4F_+^2(u+v) \rmd u\rmd v\nonumber\\
&& +\frac{F_-(u+v)}{F_+(u+v)}\rmd x^2+\cos^2(u-v)F_+^2(u+v)\rmd y^2\,.
\end{eqnarray}

The interaction region corresponds to the triangular region in the $u - v$ plane bounded by the lines $u=0$, $v=0$ and $u+v=\pi/2$. 
Following Khan-Penrose \cite{kha-pe}, the extension of the metric is then obtained simply by performing the substitution rules $u\,\rightarrow\,u\,\Theta(u)$ and $v\,\rightarrow\,v\,\Theta(v)$ in Eq. (\ref{met2}), using the Heaviside step function $\Theta$. 
The four spacetime regions
\beq
\begin{array}{lll}
u\geq 0\,,\quad v\geq 0\,,\quad u+v<\pi/2 &\quad \mbox{Region I} &\quad \mbox{(interaction)}
\\
0\leq u < \pi/2 \,,\quad v<0 &\quad \mbox{Region II} &\quad \mbox{($u$-wave)}
\\
u<0\,,\quad 0\leq v < \pi/2 &\quad \mbox{Region III } &\quad \mbox{($v$-wave)}
\\
u<0\,,\quad v<0 &\quad \mbox{Region IV} &\quad \mbox{(flat)}
\end{array}
\eeq
are shown in Fig.~\ref{fig:1}. 
The resulting extended metric is 
\begin{eqnarray}
\label{metall}
\rmd s_{II}^2&=&-4F_+^2(u) \rmd u\rmd v+\frac{F_-(u)}{F_+(u)}\rmd x^2+\cos^2u F_+^2(u)\rmd y^2\,, \nonumber\\
\rmd s_{III}^2&=&-4F_+^2(v) \rmd u\rmd v+\frac{F_-(v)}{F_+(v)}\rmd x^2+\cos^2v F_+^2(v)^2\rmd y^2\,, \nonumber\\
\rmd s_{IV}^2&=&-4\rmd u\rmd v+\rmd x^2+\rmd y^2\,.
\end{eqnarray}
In this way the extended metric in general is $C^0$ (but not $C^1$) along the null boundaries $u=0$ and $v=0$, so that the Riemann tensor acquires distributional parts. 
At the boundaries the Weyl tensor has $\delta$-functions, otherwise it is regular \cite{do-ve1} (the Ricci tensor is vanishing everywhere). 
One wave will be a function of $u$ only (in Region II), the other a function of $v$ only (in Region III).
In the $t - z$ plane Region II corresponds to the region $t-\pi\leq z\leq t$, and Region III to the region $-t\leq z\leq \pi-t$.
It is worth noting that certain calculations are more easily done in one or the other of these two sets of coordinates, so we will switch back and forth between them as needed.


\subsection{The effective optical medium}

Let us apply the optical medium analogy developed in Section 2 to the Ferrari-Iba\~nez metric (\ref{met1}).
The associated effective electric and magnetic permeability tensors (\ref{epsdef}) are given by
\begin{eqnarray}
\epsilon_{ab} &=& \mu_{ab} = {\rm diag}\left[\epsilon_1,\epsilon_2,\epsilon_3\right]\,,
\end{eqnarray}
with 
\beq
\epsilon_1=\frac1{\epsilon_2}=\frac{\cos^2z}{\epsilon_3}=\cos z\sqrt{\frac{F_+^3(t)}{F_-(t)}}\,.
\eeq
The general definition of the refraction index (\ref{refrac}) then yields
\beq
\label{nfin}
n = \sqrt{\frac{\epsilon_3}{\epsilon_1e^{2}_{1} + \epsilon_2e^{2}_{2} + \epsilon_3e^{2}_{3}}}\,,
\eeq
where $e_a$ are the components of the spatial unit vector ${\mathbf e}$ of the photon direction (see Eq. (\ref{fresneleq})).

The above relation (\ref{nfin}) can be further specified for the cases of photons propagating along the $x$-axis (${\mathbf e}=(1,0,0)$, $ n_x=\sqrt{\epsilon_2\epsilon_3}=\sqrt{F_-(t)/F_+^3(t)}$), the $y$-axis (${\mathbf e}=(0,1,0)$, $ n_y=\sqrt{\epsilon_1\epsilon_3}=\cos z$) and the $z$-axis (${\mathbf e}=(0,0,1)$, $ n_z=1$), showing then the anisotropic properties of the equivalent medium associated with the background metric: it is homogeneous with a time-dependent refraction index along the $x-$axis, while along the $y-$axis it is inhomogeneous. 

In the collision region, for fixed values of $t$ and $z$ it is $n_x < 1$ if $\sigma=1$ and $n_x > 1$ if $\sigma=-1$, whereas $n_y < 1$ for both choices of $\sigma$. Hence it follows that for a singularity-developing metric ($\sigma=-1$) the $x-$ and $y-$axes are naturally defined as the subluminal and superluminal directions of light propagation, respectively. 
In particular, the refraction index $n_x$ diverges while approaching the singularity, so that the associated effective medium becomes increasingly dense in this limit.
In the case where the solution has a horizon ($\sigma=1$), both axes are instead associated with superluminal light propagation, and the refraction index $n_x$ for photons propagating along the $x$-axis goes to zero as the horizon is approached.

As shown in Eq. (\ref{metall}), the single wave regions II and III are described by the same line element as in Eq. (\ref{met2}), but with metric components depending only on either $u$ or $v$, respectively. 
Therefore, we find
\beq
n_x=\sqrt{\frac{F_-(u)}{F_+^3(u)}}\,,\quad
n_y=\cos u\,,\quad
n_z=1\,,
\eeq
and
\beq
n_x=\sqrt{\frac{F_-(v)}{F_+^3(v)}}\,,\quad
n_y=\cos v\,,\quad
n_z=1\,,
\eeq
respectively, with $u$ and $v$ related to $t$ and $z$ by Eq. (\ref{eq:uv_trasf}). 
The equivalent media in the two single wave regions are thus inhomogeneous and have a time-dependent refraction index along both $x-$ and $y-$axis \cite{gwem_epl}.

Finally, in the flat spacetime region IV we have simply $\epsilon_1 = \epsilon_2 = \epsilon_3 = 1$, 
so that $n = n_x = n_y = n_z = 1$. 

\begin{figure}
\begin{center}
$\begin{array}{cc}
\includegraphics[scale=0.27]{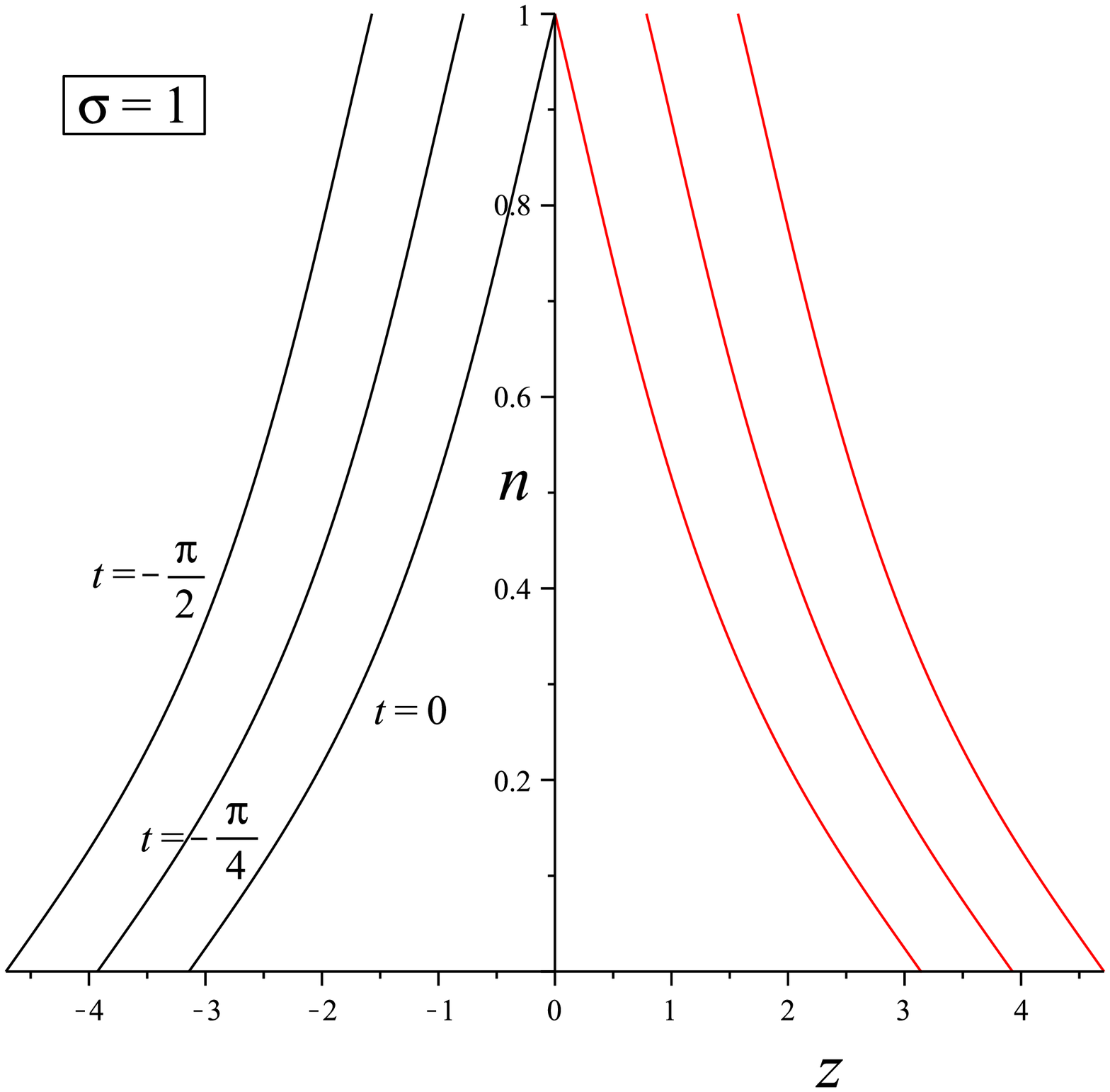}&\qquad
\includegraphics[scale=0.27]{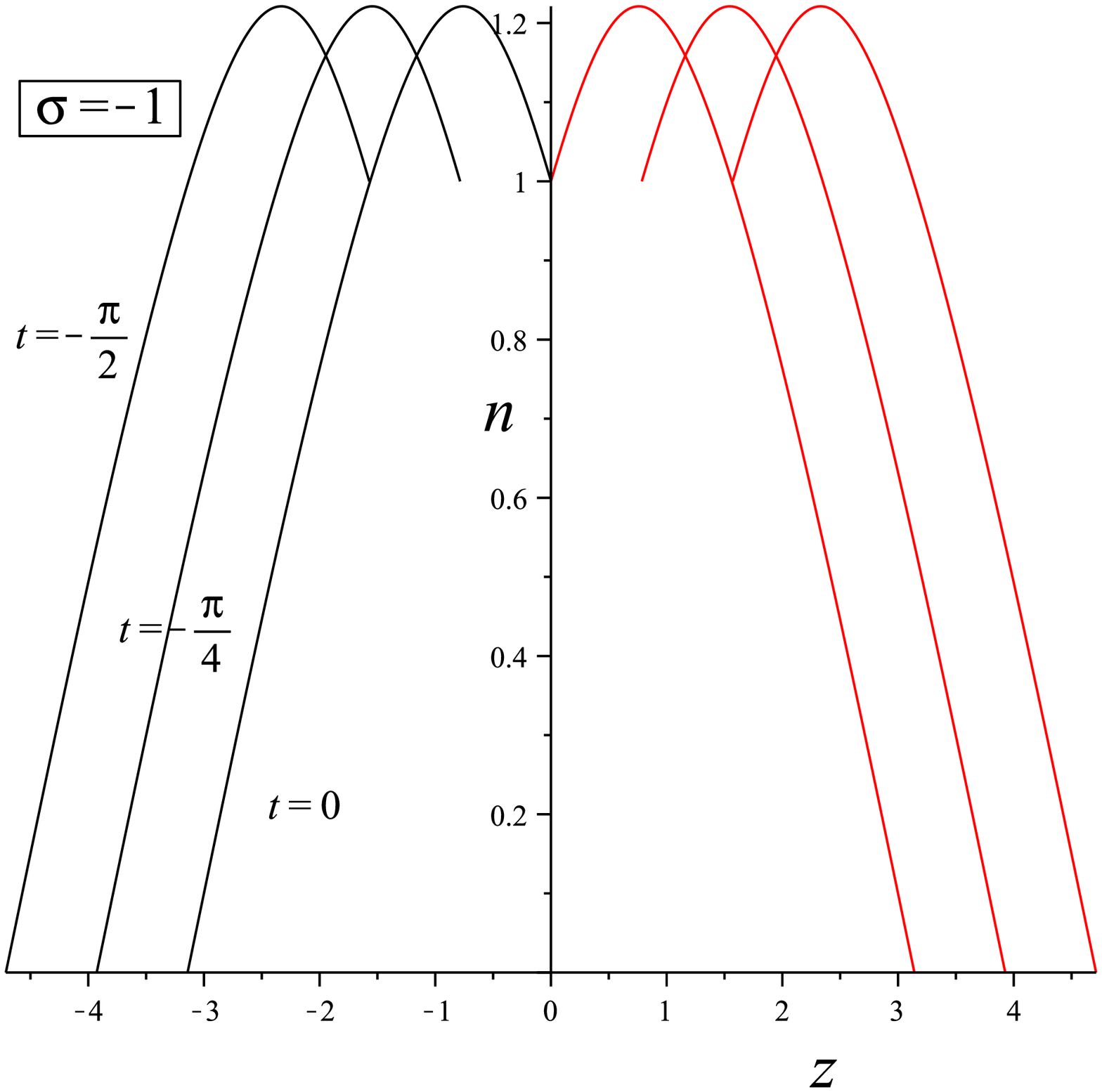}\\[.2cm]
\mbox{(a)} &\qquad \mbox{(b)}\cr
\end{array}
$\\
\end{center}
\caption{The behavior of the refraction index (\ref{nfin}) is shown as a function of $z$ for a given direction on the wave front $e_1=1/\sqrt{2}=e_2$, $e_3=0$ for different values of $t=[-\pi/2,-\pi/4,0]$ before collision (and at the time of collision $t=0$) in both cases $\sigma=\pm1$.
}
\label{fig:2}
\end{figure}

\begin{figure}
\begin{center}
$\begin{array}{cc}
\includegraphics[scale=0.27]{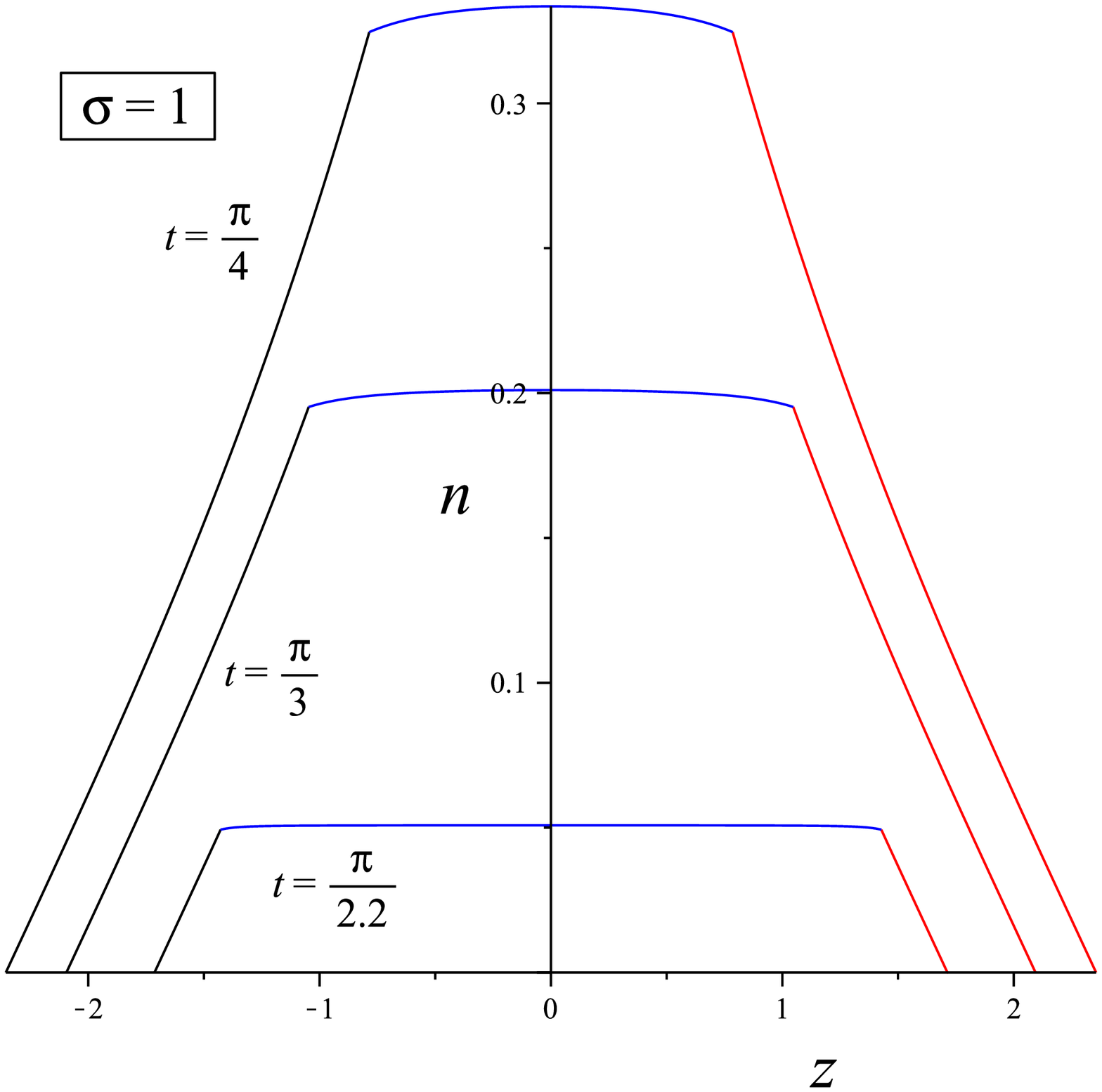}&\qquad
\includegraphics[scale=0.27]{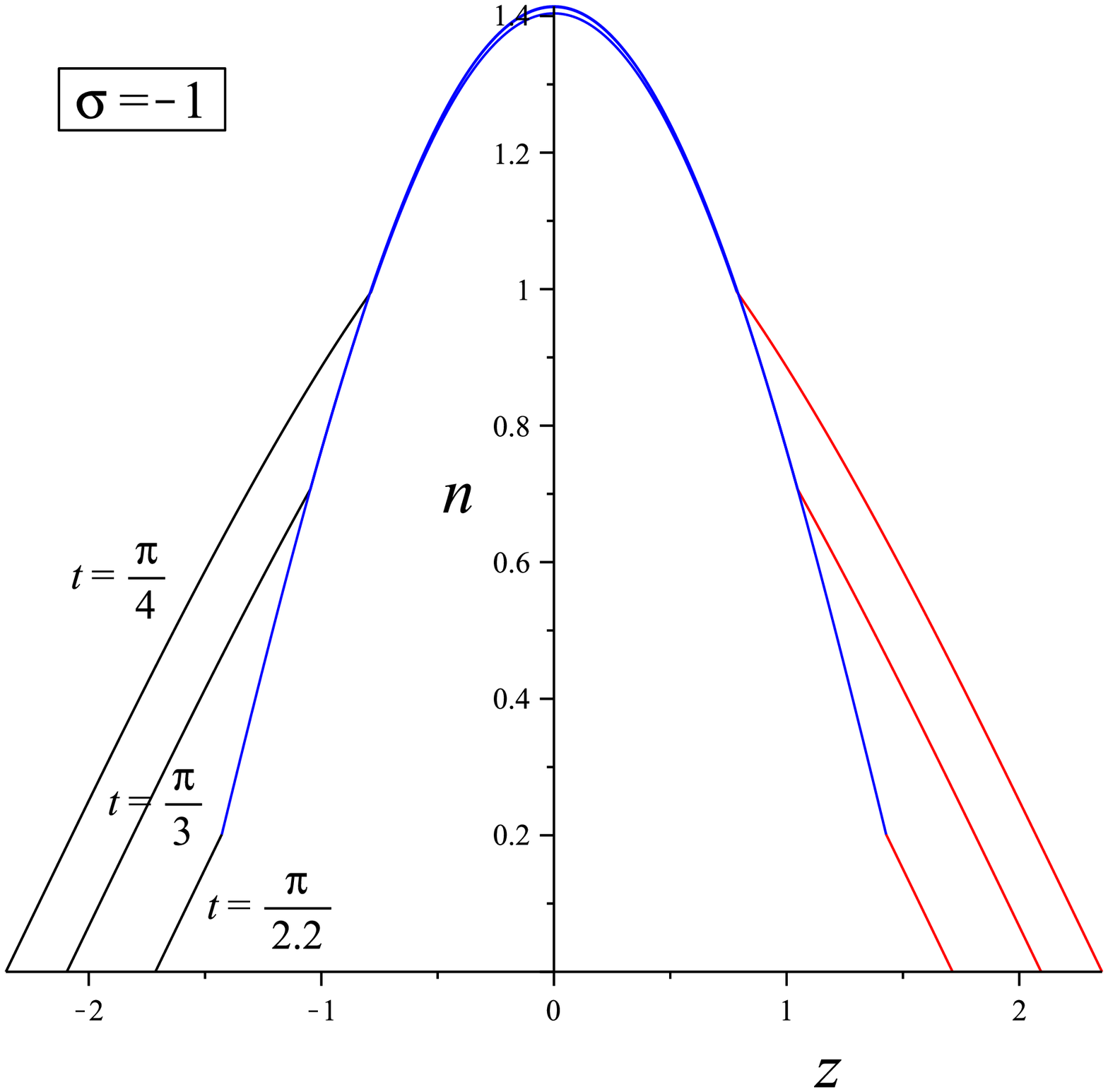}\\[.2cm]
\mbox{(a)} &\qquad \mbox{(b)}\cr
\end{array}
$\\
\end{center}
\caption{The behavior of the refraction index (\ref{nfin}) is shown as a function of $z$ for a given direction on the wave front $e_1=1/\sqrt{2}=e_2$, $e_3=0$ for different values of $t=[\pi/4,\pi/3,\pi/2.2]$ after collision in both cases $\sigma=\pm1$.
}
\label{fig:3}
\end{figure}

It is also interesting to study the behavior of the effective refraction index (\ref{nfin}) as a function of $z$ for a given direction on the wave front, i.e., $e_3=0$, and selected values of $t$ before and after the collision.
At a fixed time $t<0$, the two waves are confined in the two regions $t-\pi\leq z\leq t$ (progressive wave) and $-t\leq z\leq \pi-t$ (regressive wave), respectively.
The spacetime is flat for $-t\leq z\leq t$.
At a fixed time $t>0$, for $t-\pi\leq z\leq -t$ there is the still incoming first wave, whereas for $t\leq z\leq \pi-t$ there is the still incoming second wave. For $-t\leq z\leq t$ the two waves interact.
As the time $t$ tends to $\pi/2$, the region of interaction expands and reaches its maximum for $t=\pi/2$ when the singularity is
created between $-\pi/2\leq z\leq \pi/2$.
The above situation is summarized and illustrated in Figs. \ref{fig:2} and \ref{fig:3}.
The photon direction on the wave front has been chosen as $e_1=1/\sqrt{2}=e_2$, so that the refraction index (\ref{nfin}) becomes
\beq
n 
= \sqrt{\frac{2\epsilon_2\epsilon_3}{1+ \epsilon_2}}=n_xn_y \sqrt{\frac{2}{n_x^2+n_y^2}}\,,
\eeq
or, equivalently, 
\beq
\frac{1}{n^2}=\frac12 \left(\frac{1}{n_x^2}+\frac{1}{n_y^2}  \right)\,.
\eeq
It turns out that in the case of solutions with horizon $n$ assumes practically constant values all over the interaction region after collision.


\section{Light rays}

In the limit of geometric optics electromagnetic waves propagate along the null geodesics of the background spacetime \cite{mas87}.
In this section we study how an incoming light ray propagating along a straight line in the flat spacetime region IV is deflected when passing through the single wave region II (or to its symmetric counterpart, Region III) and finally enters the collision region I before reaching the singularity.
The solutions for the photon 4-momentum components $P^\alpha=\rmd x^\alpha/\rmd\lambda$ in each spacetime region are listed in the Appendix.
Matching conditions at the boundaries must be imposed properly, so that the photon 4-momentum be continuous everywhere.
We limit our analysis to the $t - z$ (or, equivalently, $u - v$) plane, where the most important differences between the horizon-forming and singularity-developing solutions appear, especially concerning the behavior of the geodesics approaching the hypersurface $t = \pi/2$ (or $u + v = \pi/2$).

In the flat spacetime region IV as well as in the single wave regions II and III the photon 4-momentum can be parametrized in terms of the Killing quantities $p_v$, $p_x$ and $p_y$.
In the collision region I, instead, only the specific momenta $p_x$ and $p_y$ are conserved, but there is a further constant of motion, $K$, which can be used as a parameter. 
The features of motion turn out to be very different depending on whether the 4-momentum $P_{IV}$ of the incoming photon has a nonvanishing component along the $y-$direction or not.

Let us consider first the case $p_y=0$.
At the boundary II-I where $t=u$ (hereafter, we suppress the dependence of the various functions on $u$ to simplify the notation) we have
\begin{eqnarray}
\label{confronto}
P_{II}&=& -\frac{p_v}{2F_+^2}\left(\partial_u +\frac{p_x^2}{n_x^2 p_v^2} \partial_v \right)+\frac{p_x}{n_x^2F_+^2}\partial_x\,,\nonumber\\
P_{I}&=& \frac{1}{2F_+^2}\left(\sqrt{\frac{p_x^2}{n_x^2}+K^2}-K \right)\partial_u\nonumber\\
&& +\frac{1}{2F_+^2}\left(\sqrt{\frac{p_x^2}{n_x^2}+K^2}+K \right)\partial_v
 +\frac{p_x}{n_x^2F_+^2}\partial_x\,,
\end{eqnarray}
where the refraction index along the $x$-axis, $n_x$,  follows from  Eq. (\ref{nfin}) with ${\mathbf e}=(1,0,0)$ and it is such  that $F_-=n_x^2 F_+^3$.
Note that here we use $\pm \sqrt{K^2}=K$, permitting both sign choices of $K$. 
It is easy to recognize that the $x-$components already agree, whereas continuity for the $u-$ and $v-$components implies
\beq
-p_v=\sqrt{\frac{p_x^2}{n_x^2}+K^2}-K\,,\qquad
-\frac{p_x^2}{p_v n_x^2}= \sqrt{\frac{p_x^2}{n_x^2}+K^2}+K\,,
\eeq
and both equations are satisfied by 
\beq
\label{Kcond}
K=\frac{p_v}{2}-\frac{p_x^2}{2p_vn_x^2}\,.
\eeq
The previous equation then yields the proper value of $K$ allowing for the geodesic path to be continued in Region I.
Integrating the geodesic equations in Region II for a given set of initial conditions and selected values of the constants $p_v$ and $p_x$ identifies the value of $u$ at which the photon enters the interaction region.

\begin{figure}
\begin{center}
$\begin{array}{cc}
\includegraphics[scale=0.3]{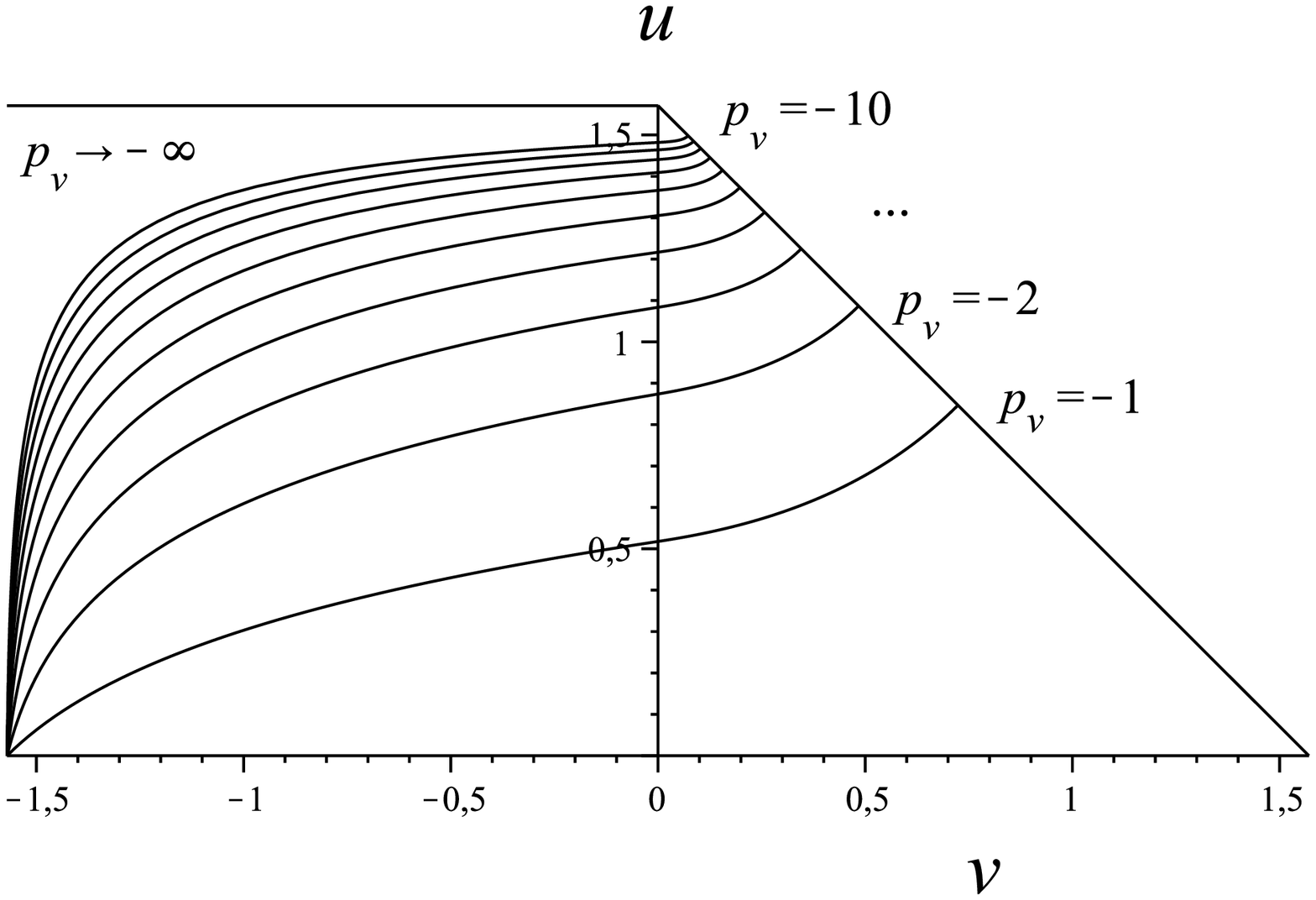}&\qquad
\includegraphics[scale=0.3]{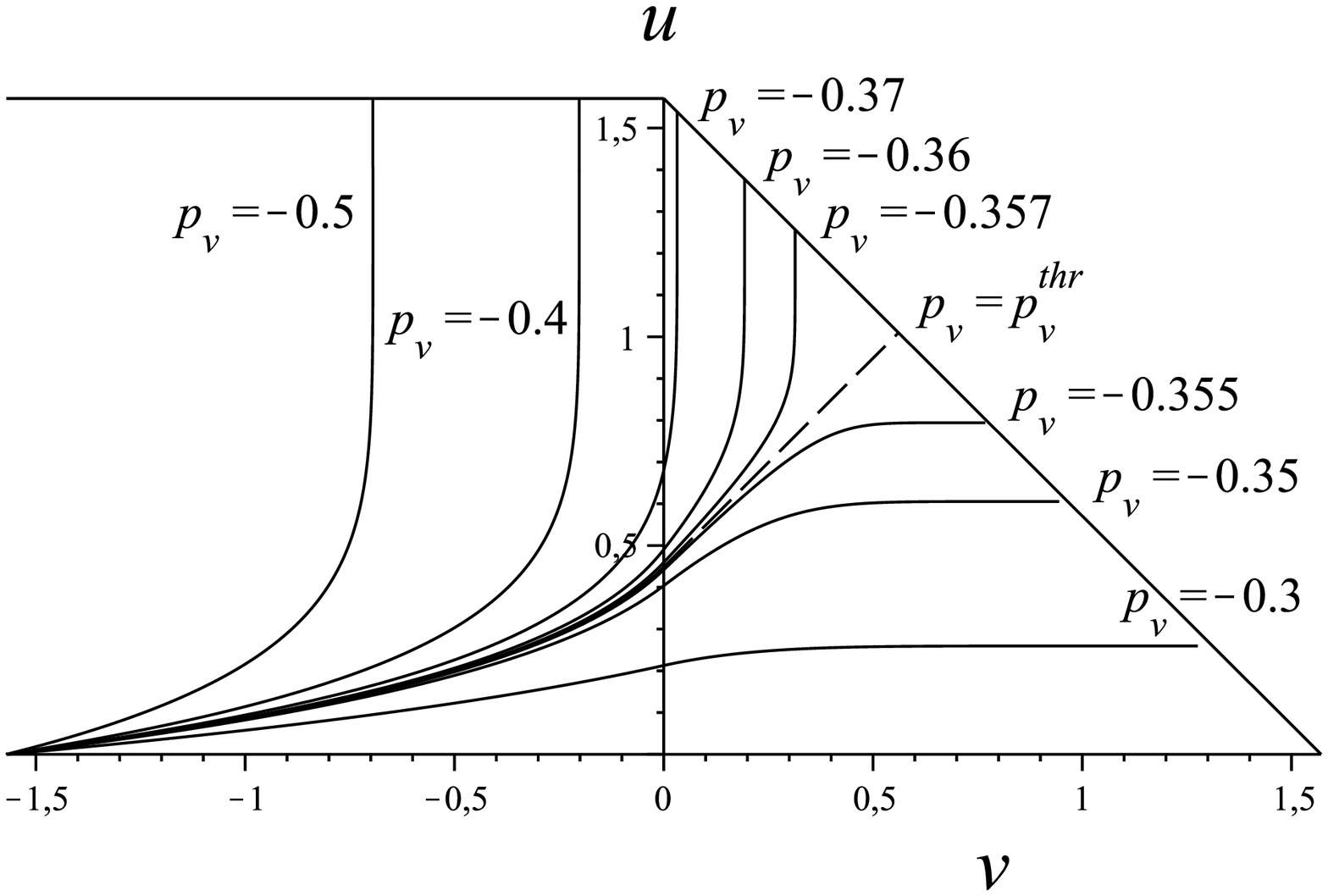}\\[.2cm]
\mbox{(a)} &\qquad \mbox{(b)}\cr
\end{array}
$\\
\end{center}
\caption{The features of geodesic motion in the $u - v$ plane are shown in panels (a) and (b) in the case $\sigma = 1$ (horizon) and $\sigma =-1$ (singularity), respectively.
The geodesics equations ${\rmd u}/{\rmd \lambda}=P^u$ and ${\rmd v}/{\rmd \lambda}=P^v$ have been integrated in the single $u$-wave region II with initial conditions $u(0)=0$, $v(0)=-\pi/2$ and the choice of parameters $p_x = 1$, $p_y = 0$.
The curves correspond to different values of $p_v$.
The numerical integration of the orbits has then been continued in the collision region I by imposing the matching at the boundary II-I, i.e., $v=0$.
For $\sigma = 1$ the point ($u=\pi/2$, $v=0$) represents an accumulation point for the null geodesics leading to a fold singularity. 
The critical value of $p_v$ discriminating among the two kinds of orbits in the case $\sigma =-1$ is $p_v^{\rm thr}\approx-0.35578$.
}
\label{fig:4}
\end{figure}

An example for the integration of the orbits for $p_y=0$ is shown in Fig. \ref{fig:4}.
The behavior of null geodesics turns out to be different depending on whether the solution is horizon-forming or singularity-developing.
In the former case the motion has a global homogeneity for all values of $p_v$. 
Only positive values of $K$ are allowed.
The limit $p_v\to-\infty$ gives the so-called fold singularities at the points of the horizon $(u=\pi/2,v=0)$ and $(u=0,v=\pi/2)$, already discussed by Dorca and Verdaguer \cite{do-ve1}.
The presence of such singularities is the origin of an accumulation of the null geodesics on the surface $v=0$ for $u\to\pi/2$, as shown in Fig. \ref{fig:4} (a).
As follows from the analysis of the refraction index, the geodesics approach the horizon always perpendicularly.
In fact, let us examine the slope of the trajectories
\beq
\label{slope}
\frac{\rmd u}{\rmd v}=\frac{\sqrt{\frac{p_x^2}{n_x^2}+K^2}-K}{\sqrt{\frac{p_x^2}{n_x^2}+K^2}+K}\,,
\eeq
by using Eq. (\ref{confronto}), as the horizon is approached.
In this limit, i.e., $t=u+v\to\pi/2$, the refraction index behaves as $n_x\to 0$, implying that ${\rmd u}/{\rmd v}\big|_{t\to \pi/2} \to 1$.

In the case of solutions with a singularity the geodesics exhibit a typical twofold behavior for $p_y=0$ (see Fig. \ref{fig:4} (b)). This is due to the existence of a critical value of $p_v$ in the single wave region, implying $K=0$ for a given choice of $p_x$ as well as initial conditions. We thus identify those geodesics that propagate through the collision region at a constant value of $z$, irrespective of the presence of the waves, i.e., $p_v^{\rm thr}=-{|p_x|}/{n_x}$ (see Eq. (\ref{Kcond})).
In contrast to the case of the horizon-forming solution, every value of $K$ is now allowed.
The null geodesics approach the singularity in a way which depends on the sign of $K$.
In fact, since here the refraction index $n_x\to \infty$, the slope of the trajectories (\ref{slope}) in the limit $t\to\pi/2$ is ${\rmd u}/{\rmd v}\big|_{t\to \pi/2} \to {(|K|-K)}/{(|K|+K)}$.
Therefore, if $K>0$ (i.e., the value of $p_v$ of the incoming photon is greater than the threshold value $p_v^{\rm thr}$), the slope turns out to be ${\rmd u}/{\rmd v}\big|_{t\to \pi/2}\to 0$ (the orbit is parallel to the $v-$axis). On the other hand, if $K<0$ (i.e., $p_v<p_v^{\rm thr}$) the slope grows indefinitely, namely ${\rmd u}/{\rmd v}\big|_{t\to \pi/2}\to \infty$ (the orbit is parallel to the $u-$axis).
There also exists a limiting value of $p_v$ below which the photon does not enter the interaction region at all.
The case $K=0$ represents a separatrix, in the sense that in this case the orbit will approach the singularity perpendicularly at a fixed value of $z$, as discussed above.


\begin{figure}
\begin{center}
$\begin{array}{cc}
\includegraphics[scale=0.3]{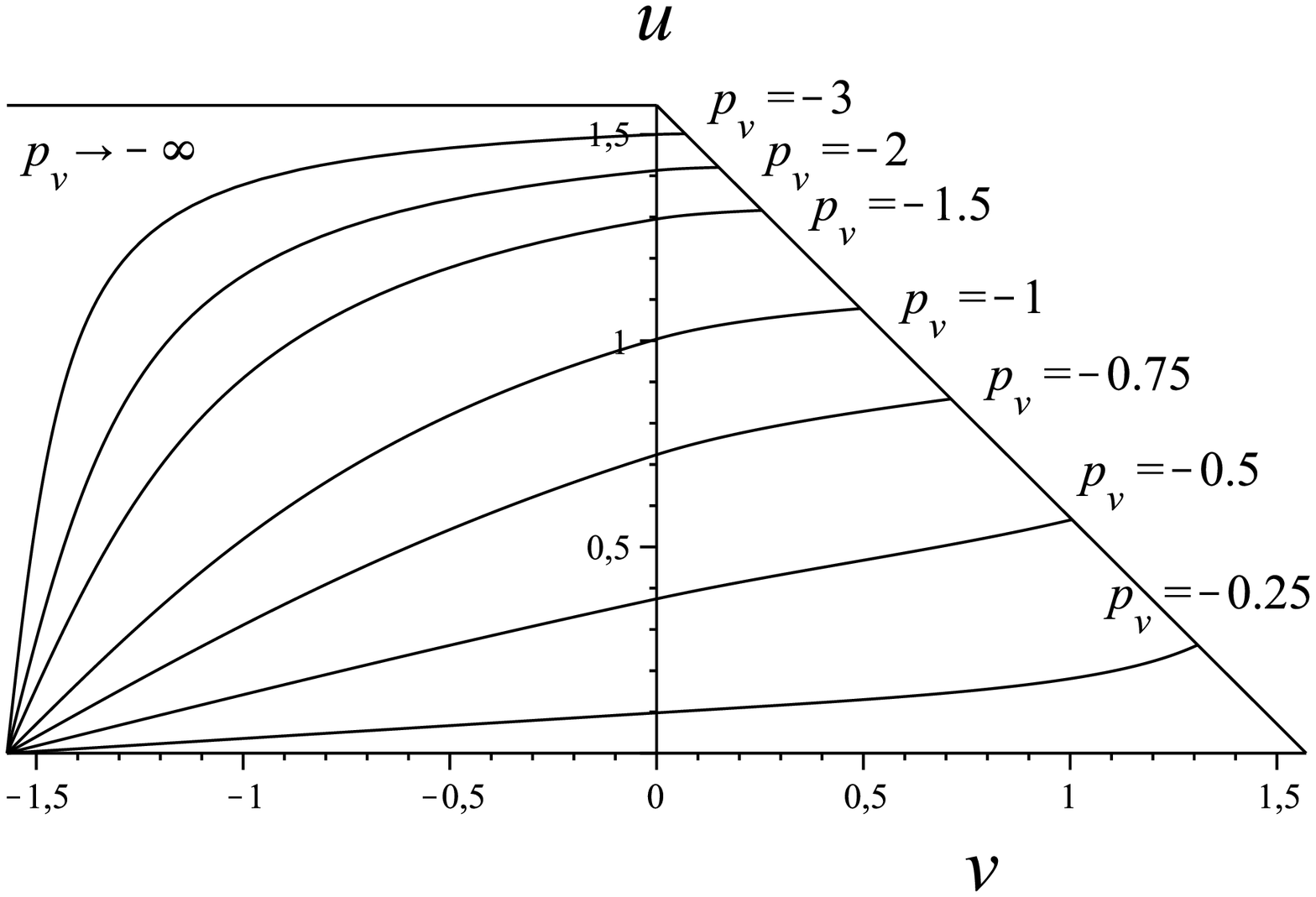}&\qquad
\includegraphics[scale=0.3]{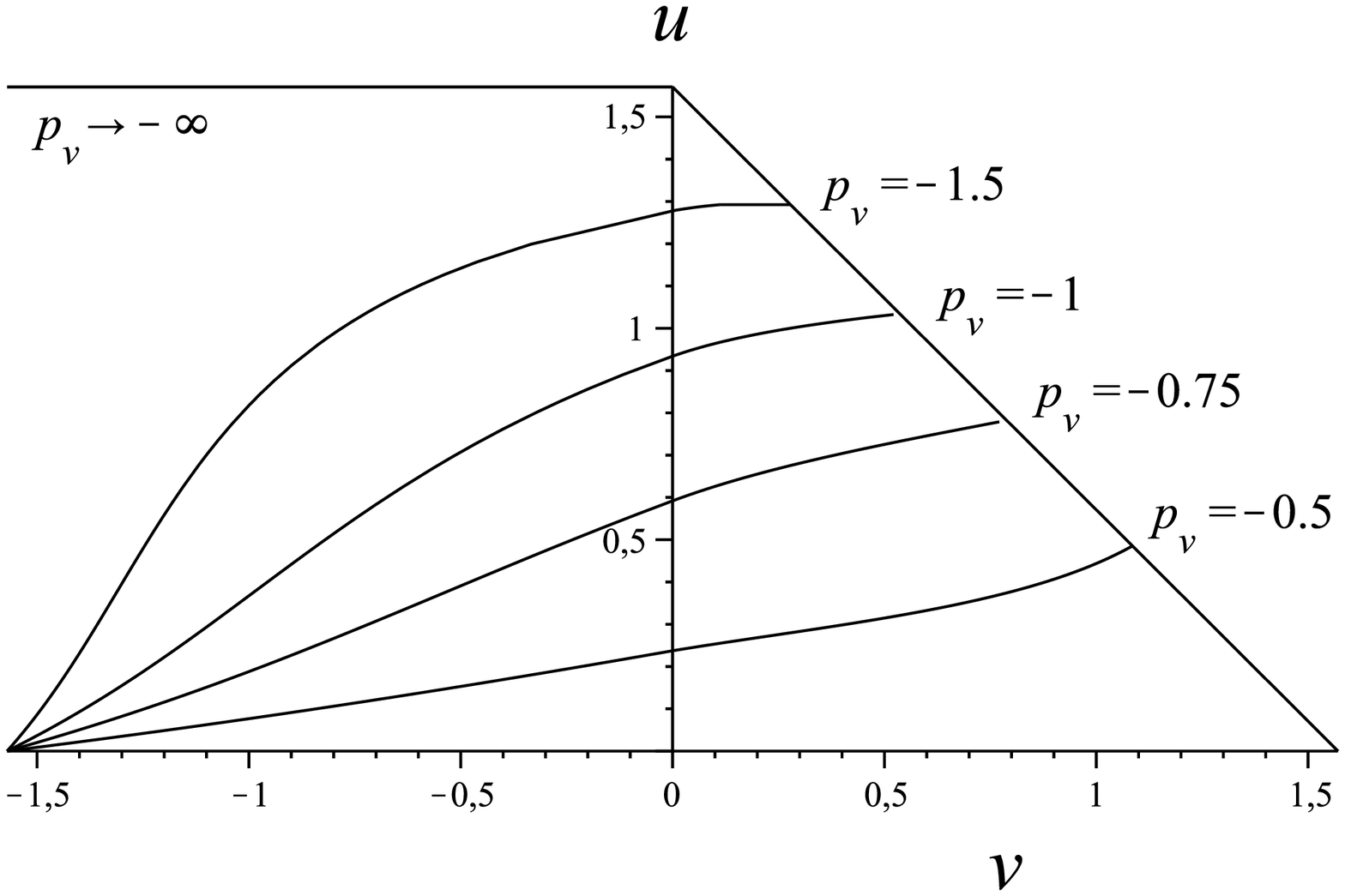}\\[.2cm]
\mbox{(a)} &\qquad \mbox{(b)}\cr
\end{array}
$\\
\end{center}
\caption{The typical behavior of null geodesics with $p_y\not=0$ in shown for both kinds of solutions ($\sigma = \pm1$).
The choice of initial conditions in the single $u-$wave region II is the same as in Fig. \ref{fig:4}, whereas the  parameters $p_x$ and $p_y$ have been set to $p_x = 0$, $p_y = 1$ in the case of a horizon-forming metric shown in panel (a) and to $p_x = 1$, $p_y = 1$ in the case of a singularity-developing metric shown in panel (b).
The curves correspond to different values of $p_v$.
The fold singularity at the point ($u=\pi/2$, $v=0$) arises for $p_v\to-\infty$ in both cases.
}
\label{fig:5}
\end{figure}

A similar discussion can be made in the case of nonvanishing $p_y\not=0$ for the incoming photon.
The continuity conditions at the boundary II-I for the momentum components now imply
\begin{eqnarray}
\label{matchpy}
-p_v&=&\sqrt{\frac{p_x^2}{n_x^2}+p_y^2+K^2}\mp\sqrt{K^2+p_y^2-\frac{p_y^2}{n_y^2}}\,,\nonumber\\
-\frac1{p_v}\left(\frac{p_x^2}{n_x^2}+\frac{p_y^2}{n_y^2}\right)&=&\sqrt{\frac{p_x^2}{n_x^2}+p_y^2+K^2}\pm\sqrt{K^2+p_y^2-\frac{p_y^2}{n_y^2}}\,,
\end{eqnarray}
where $n_y=\cos z=\cos u$, as given by Eq. (\ref{nfin}) with ${\mathbf e}=(0,1,0)$.
Solving the system of equations (\ref{matchpy}) for $K^2$ gives 
\beq
K^2=\frac1{4p_v^2}\left(p_v^2+\frac{p_x^2}{n_x^2}+\frac{p_y^2}{n_y^2}\right)^2-\frac{p_x^2}{n_x^2}-p_y^2\,.
\eeq
Back-substituting into Eq. (\ref{matchpy}) implies that the upper sign only is allowed in order to fulfill the matching conditions for both cases $\sigma=\pm1$, i.e., if $p_y\not=0$ only those orbits of photons having a 4-momentum (\ref{geo_null_interaction}) with a positive $z-$component can be continued across the boundary II-I.
The threshold disappears and the geodesic motion exhibits a common behavior for both 
horizon-forming and singularity-developing solutions, including the way to reach the hypersurface $t=\pi/2$. 
This peculiar asymmetry characterizing the behavior of null geodesics depending on the presence/absence of a $y-$component in the incoming 4-momentum can be associated with a different role played by the $x-$ and $y-$coordinates spanning the surface of the wavefront in these spacetimes, as discussed in Ref. \cite{bcl} in the case of timelike geodesics.
An example of numerical integration of the orbits with $p_y\not=0$ is shown in Fig. \ref{fig:5}.


\section{Closed rectangular paths on the wave front}

Since we have chosen the wave to propagate along the $z-$axis, optical effects in that direction turn out to be trivial, as is evident also from the resulting effective refraction index along such a direction, $n_z=1$. We want to investigate the optical properties of the equivalent medium on the wave front, i.e. on the $x-y$ plane.
To this end, let us imagine a square optical path, centered at the origin of the $x-y$ plane, and let $L$ be the length of each side, obtained for instance by using optical guides.
Photons moving along the $x-$direction will travel the time $T_x$ along each side of the square which is parallel to the $x-$axis, such that
\beq
\label{tx}
\int^{T_x}_0 \frac{dt}{n_x(t)}=L
\eeq
and the time $T_y$
\beq
T_y=L n_y(z)=L\cos z
\eeq
along each side of the square which is parallel to the $y-$axis. 
Solving for $T_x$, Eq. (\ref{tx}) then gives
\beq
T_x=\sigma \arcsin \left[2W_0\left(-\frac12 e^{-\frac{1+\sigma L}{2}}\right)+1\right]\,,
\eeq
where $W_0(\xi)$ denotes the principal branch of the Lambert $W$ function\footnote{
The Lambert $W$ function has an infinite number of solutions for each (non-zero) value of $\xi$, i.e.,  it  has an infinite number of branches.
Let us denote by $W_k(\xi)$ the $k-$branch, where $k$ is any non-zero integer.
If the variable $\xi$ is real, then there are two possible real values of $W(\xi)$ in the interval $-1/e\leq \xi <0$.
The branch satisfying $W(\xi)\geq -1$ is denoted by $W_0(\xi)$ and is referred to as the principal branch of the $W$ function, whereas the branch satisfying $W(\xi)\leq-1 $ is denoted by $W_{-1}(\xi)$.
}, i.e. the special function satisfying the equation $W(\xi)e^{W(\xi)}=\xi$ \cite{corless}.

\begin{figure}
\begin{center}
\includegraphics[scale=0.35]{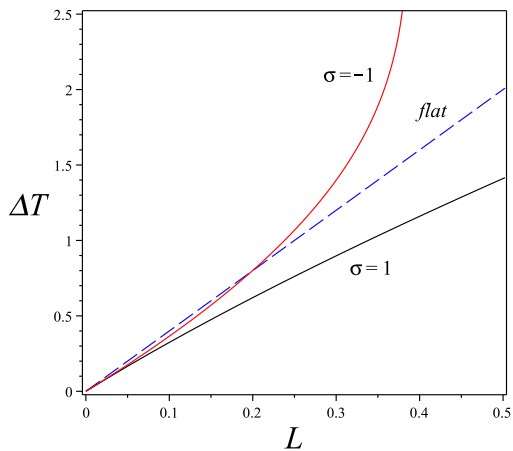}
\end{center}
\caption{The elapsed time corresponding to photons making a square path on the wave front in the collision region (for horizon-forming as well as singularity-developing colliding gravitational waves) is plotted as a function of the length of the square side $L$. Here we have set $z=\pi/4$, but the behavior is similar for a different choice of $z$.
The case of a flat spacetime is also shown for comparison.
}
\label{fig:6}
\end{figure}

A full path will then correspond to the elapsed time
\beq
\Delta T=2T_x+2T_y\,.
\eeq
Its behavior as a function of $L$ for a fixed value of $z$ is shown in Fig. \ref{fig:6}. 
The explicit value of $\Delta T$ strongly depends on the spacetime region where it is evaluated.
In the flat spacetime, one would simply have
\beq
\label{deltatflat}
\Delta T_{\rm (flat)}=4L\,.
\eeq
Therefore, apart from the manifest difference from the flat spacetime case, measuring $\Delta T$ would also allow to clearly distinguish among the horizon-forming and the singularity-developing metrics.


\section{Concluding remarks}

We have studied light propagation in a colliding gravitational wave spacetime by using the optical medium analogy.
According to such an approach, the description of electromagnetic fields in a curved spacetime is equivalently accomplished by solving Maxwell's equations in a flat spacetime but in the presence of a medium, whose properties are fully specified by the associated constitutive relations.
We have considered as a background solution the Ferrari-Iba\~nez metrics, in which the interaction region of the colliding waves propagating along the common $z-$direction is of Petrov type-D, with either a Killing-Cauchy horizon or a singularity formed at a finite distance from the collision plane, after a finite time from the instant of collision.
The most relevant physics involves the plane transverse to the direction of propagation of the waves.
The effective medium turns out to be homogeneous and with an associated time-dependent refraction index along the $x-$axis, whereas along the $y-$axis it is inhomogeneous.
Moreover, the coordinate components of the photon spatial velocity become greater than 1 along the $y-$axis for metrics of both kinds, while the $x-$axis is a superluminal direction of light propagation only for solutions with a horizon. 
These optical properties affect for instance the time a photon spends inside an optical guide to travel along a given path, whose measurement by a simple interferometric device would provide information about the nature of the spacetime region where it is evaluated.

The possibility of a photon's coordinate speed becoming greater than the speed of light in vacuum is a very interesting feature which may also have a counterpart in experiments. 
In fact, current technologies allow for the construction of the so-called ``metamaterials'' \cite{liu}, with refraction indices that are, in general, time-dependent and may be very low (less than 1) or even negative.
Metamaterial formulations for several nontrivial curved spacetime scenarios have been developed in recent years, including the Schwarzschild, Schwarzschild-de Sitter, Kerr and Kerr-Newman spacetimes, which are associated with negative phase-velocity propagation of light (see, e.g., Ref. \cite{mackay} and references therein).
Therefore, in principle, one can also arrange for an analogue material exhibiting the optical properties of the colliding gravitational wave spacetime considered here and compare the geometrization of physical interactions (otherwise impractical to explore) with experimental data by laboratory-based simulations.

\begin{acknowledgements}
We are indebted to Dr. A. Ortolan for useful discussions.
\end{acknowledgements}

\appendix

\section{Null geodesics}

We list below the solutions for the photon 4-momentum in each spacetime region, i.e., inbound flat, single wave and collision region.

The constant photon 4-momentum $P_{IV}$ of the incoming photon (Region IV) can be parametrized in terms of the conserved specific momenta $p_v$, $p_x$ and $p_y$ associated with the three Killing vectors $\partial_v$, $\partial_x$, $\partial_y$ as
\begin{eqnarray}
P_{IV}&=&-\frac{p_v}{2}\left(\partial_u+\frac{p^{2}_{\perp}}{p^{2}_{v}}\partial_v\right)+p_{\perp}\nonumber\\
&=&-\frac{p_v}{2}\left(1+\frac{p^{2}_{\perp}}{p^{2}_{v}}\right)\partial_t+p_{\perp}
-\frac{p_v}{2}\left(-1+\frac{p^{2}_{\perp}}{p^{2}_{v}}\right)\partial_z\,,
\end{eqnarray}
where $p_{\perp}=p_x\partial_x+p_y\partial_y$ (with $p_\perp^2=p_x^2+p_y^2$ using the flat spacetime notation for convenience) and $p_v < 0$ to ensure that $P_{IV}$ is future-pointing. We have made use of the relations $\partial_t =(\partial_u +\partial_v)/2$ and $\partial_z =(\partial_v -\partial_u)/2$, as follows from Eq. (\ref{eq:uv_trasf}).

In the single $u-$wave region II (and similarly in the $v-$wave region III) the conservation of $p_v$, $p_x$ and $p_y$ still holds ($\partial_v$, $\partial_x$, $\partial_y$ are Killing fields), permitting the parametrization of the photon 4-momentum as
\begin{eqnarray}
P_{II}&=&-\frac{p_v}{2F_+^2(u)}\bigg[\partial_u +\frac{1}{p^{2}_{v}}\bigg(p^{2}_{x}\frac{F_+^3(u)}{F_-(u)}+\frac{p^{2}_{y}}{\cos^2 u}\bigg)\partial_v\bigg]\nonumber\\
&&+p_x\frac{F_+(u)}{F_-(u)}\partial_x+\frac{p_{y}}{\cos^2 u F_+^2(u)}\partial_y\,.
\end{eqnarray}

In the collision region I only $\partial_x$ and $\partial_y$ are Killing vectors, implying that only the specific momenta $p_x$ and $p_y$ are conserved.
Nevertheless, there exists a further constant of motion (here denoted by $K$) related to the separation of the Hamilton-Jacobi equation \cite{do-ve1}, so that the photon 4-momentum turns out to be given by
\begin{eqnarray}
\label{geo_null_interaction}
P_{I}&=&\frac{1}{F_+^2(t)}\sqrt{p_x^2\frac{F_+^3(t)}{F_-(t)}+p_y^2+K^2}\partial_t
\pm \frac{1}{F_+^2(t)}\sqrt{K^2-p_y^2\,\tan^2 z}\partial_z\nonumber\\
&& +p_x\,\frac{F_+(t)}{F_-(t)}\partial_x +\frac{p_y}{\cos^2 z\,F_+^2(t)}\partial_y\,,
\end{eqnarray}
where the $\pm$ signs account for orbits with either increasing $(+)$ or decreasing $(-)$ values of $z$.

In order to extend a geodesic from Region II to Region I, the value of $K$ must be selected properly, so that the continuity of the $t-$ and $z-$ (or, equivalently, the $u-$ and $v-$) components of the 4-momentum is guaranteed at the boundary II-I, where $v=0$.

\end{document}